\documentclass[iop]{emulateapj}

\usepackage{natbib}
\usepackage{longtable}
\usepackage{amsmath}
\shorttitle{Virial BH Mass Estimates for 280,000 AGNs from SDSS}
\shortauthors{Koz{\l}owski}

\begin{document}

\title{Virial Black Hole Mass Estimates for 280,000 AGNs from the SDSS Broad-Band Photometry and Single Epoch Spectra}

\author{Szymon~Koz{\l}owski\altaffilmark{1}}

\altaffiltext{1}{Warsaw University Observatory, Al. Ujazdowskie 4, 00-478 Warszawa, Poland; simkoz@astrouw.edu.pl}

%%%%%%%%%%%%%%%%%%%%%  ABSTRACT  %%%%%%%%%%%%%%%%%%%%

\begin{abstract}
We use the Sloan Digital Sky Survey (SDSS) Quasar Data Release 12 (DR12Q), containing nearly 300,000 AGNs,
to calculate the monochromatic luminosities at 5100\AA, 3000\AA, and 1350\AA, derived from the broad-band extinction-corrected SDSS magnitudes.
After matching them to their counterparts based on spectra and published in the SDSS Quasar Data Release 7 (DR7Q), we find
perfect correlations with minute mean offsets ($\sim$0.01 dex) and dispersions of differences of 0.11, 0.10, 0.12 dex, respectively,
across a 2.5 dex luminosity range.
We then estimate the active galactic nuclei (AGNs) black hole masses using the broad line region radius--luminosity relations and the FWHM of the MgII and CIV emission lines,
to provide a catalog of 283,033 virial black hole mass estimates (132,451 for MgII, 213,071 for CIV, and 62,489 for both)
along with the bolometric luminosity and the Eddington ratio estimates for $0.1<z<5.5$ and for roughly a quarter of the sky covered by SDSS. The black hole mass estimates from MgII
turned out to be closely matched to the ones from DR7Q with the dispersion of differences of 0.34 dex across a $\sim$2 dex BH mass range. We uncovered a bias in the derived CIV FWHMs
from DR12Q as compared to DR7Q, that we correct empirically. 
The CIV BH mass estimates should be used with caution because the CIV line is known to cause problems in the BH mass estimation from single-epoch spectra.
Finally, after the FWHM correction, the AGN black hole mass estimates from CIV closely match the DR7Q ones
(with the dispersion of 0.28 dex), and more importantly the MgII and CIV BH masses agree internally with the mean offset of 0.07 dex and the dispersion of 0.39 dex.
\end{abstract}


%%%%%%%%%%%%%%%%%%%%%  INTRODUCTION  %%%%%%%%%%%%%%%%%%%%

\section{Introduction}

It is now widely accepted that every active galactic nucleus (AGN) hosts a supermassive black hole (BH) at its center
that accretes matter, leading to the formation of an extremely luminous accretion disk (e.g., \citealt{1973A&A....24..337S,1964ApJ...140..796S}). 
Both the monochromatic and bolometric luminosity $L$ of the disk
is tightly correlated with the radius $R$ of the broad line region (BLR), $R \propto L^{1/2}$, as shown from the reverberation mapping studies of nearby AGNs (e.g., \citealt{2000ApJ...533..631K,2002MNRAS.337..109M,2007ApJ...659..997K,2009ApJ...697..160B}). We then know the 
distance to and the the velocity of the BLR gas-dust clouds, obtained from the widths of broadened emission lines in AGN spectra,
to measure the BH mass via the virial theorem $M_{BH}\propto R v^2$ (e.g.,\citealt{2006ApJ...641..689V}; see \citealt{2013BASI...41...61S} for a review).
The luminosity of the accretion disk continuum is typically measured from the calibrated AGN spectra 
at preselected wavelengths (5100\AA, 3000\AA, and 1350\AA), however, if such a calibration is lacking it can be successfully estimated from 
the broad-band magnitudes (\citealt{2015AcA....65..251K}).

\cite{2016arXiv160806483P} provide the largest to date, uniform catalog of 297,301 AGNs with measured widths of either CIV, CIII], 
and/or MgII lines (measured as both the full width at half maximum (FWHM) 
and the blue/red half width at half maximum (HWHM)), but lacks the monochromatic luminosities, necessary to estimate the BH masses, the bolometric luminosities, and the Eddington ratios.
In this paper, we use the \cite{2015AcA....65..251K} prescription to convert the broad-band magnitudes to the monochromatic luminosities.
We then derive the ``virial BH masses'', the bolometric luminosities, and the Eddington ratios by combining the monochromatic luminosities
with the line widths via the virial theorem. We provide these quantities in a catalog form that is line-by-line matched to the 
original catalog from \cite{2016arXiv160806483P}. 

In Section~\ref{sec:methods}, we present the methodology of estimation of both the luminosity and black hole mass. 
We discuss plausible issues in Section~\ref{sec:discussion} and the paper is summarized in Section~\ref{sec:summary}.

%%%%%%%%%%%%%%%%%%%%%%%%%% DATA  %%%%%%%%%%%%%%%%%%%%

\section{Methodology}
\label{sec:methods}

We have downloaded the DR12Q data set presented in \cite{2016arXiv160806483P}. It contains 297,301 objects for which we extracted
redshifts $z$, absolute magnitudes $M_i$, FWHMs of CIV, CIII], and MgII lines (uncertainties are not provided), 
and also HWHMs of the blue and red sides for these lines, the SDSS $ugriz$ magnitudes,
and their corresponding extinctions (obtained originally from \citealt{2011ApJ...737..103S}).

\begin{figure*}
\centering
\includegraphics[width=5.5cm]{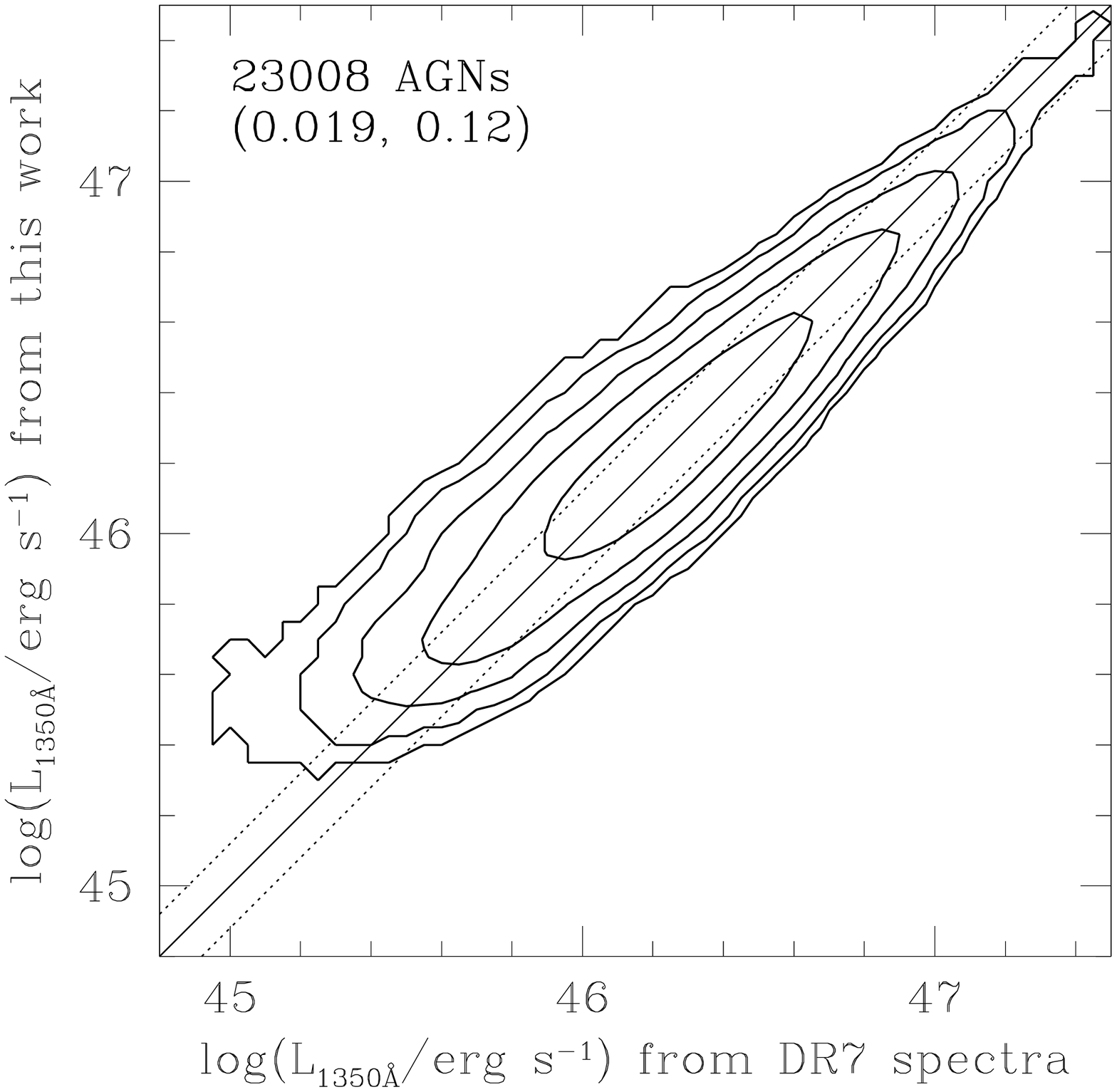}
\includegraphics[width=5.5cm]{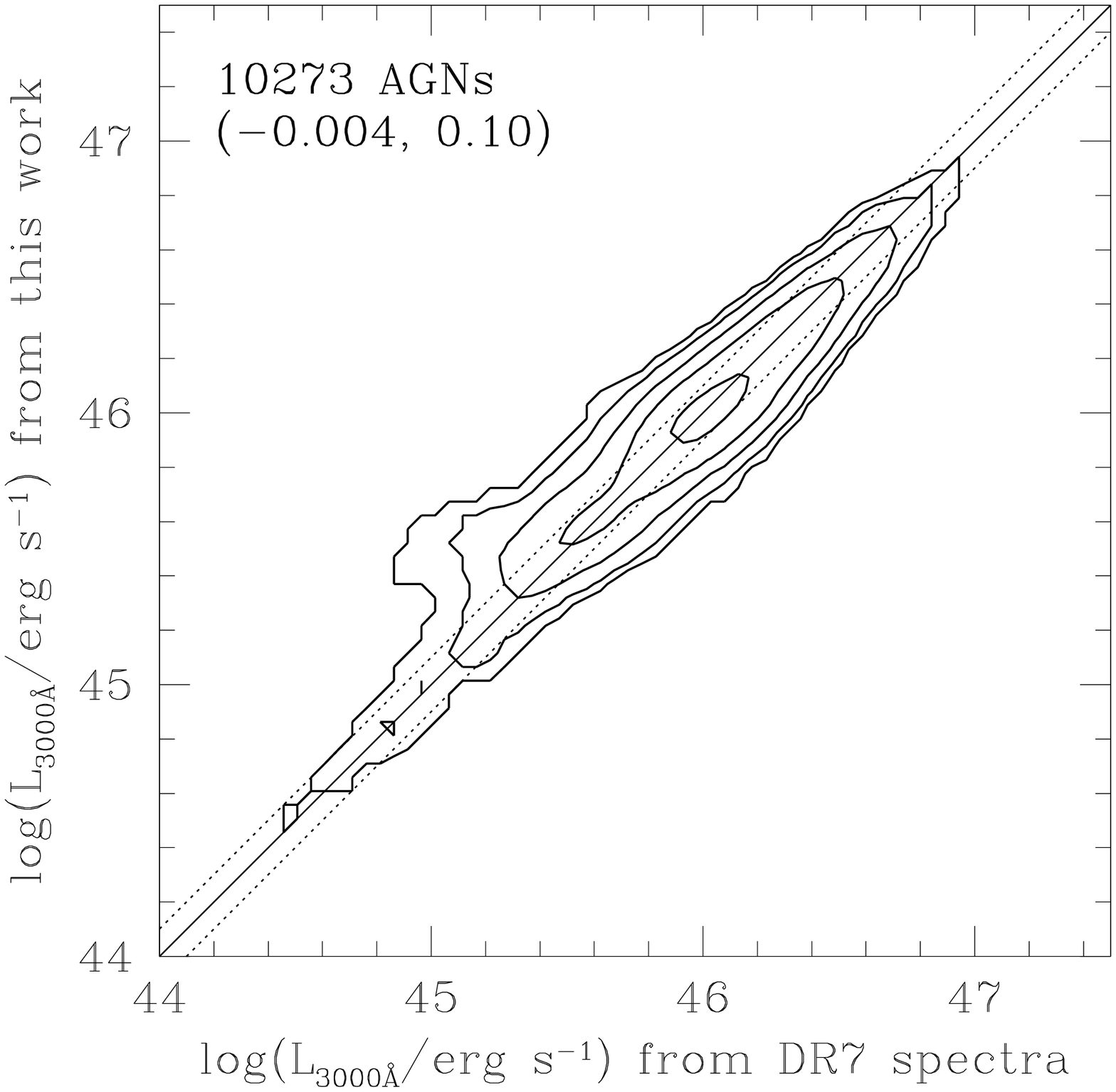}
\includegraphics[width=5.5cm]{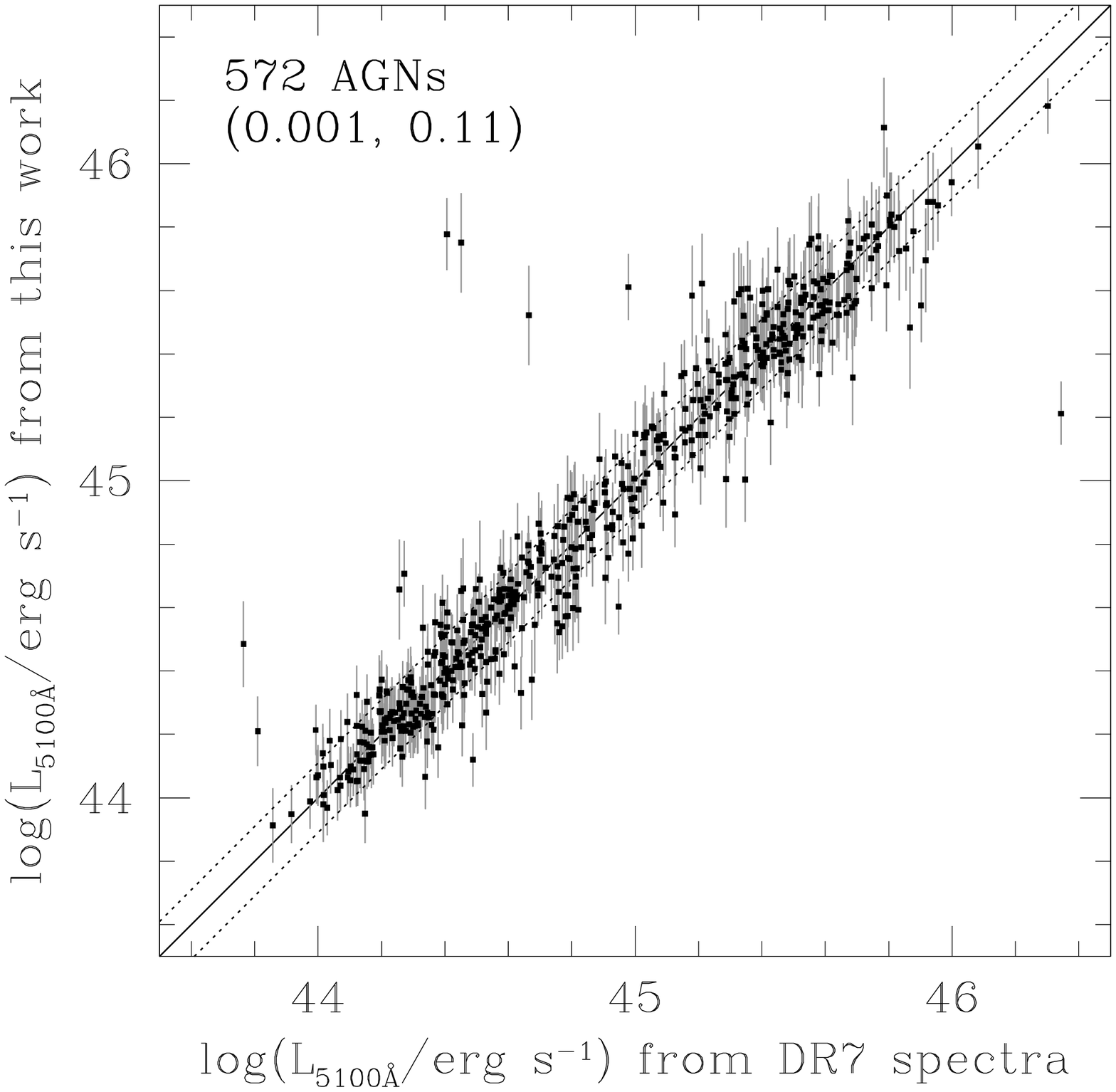}
\caption{Comparison between the luminosity at 1350\AA\ (left panel), 3000\AA\ (middle), and 5100\AA\ (right) taken from the SDSS DR7Q (x-axis) and calculated from the 
DR12Q broad-band magnitudes (y-axis). Contours show the density of objects per $0.05 \times 0.05$ dex bin. In brackets, we provide the mean and dispersion of differences, respectively.}
\label{fig:Lum}
\end{figure*}

\subsection{The Luminosity}

We converted the broad-band extinction-corrected $ugriz$ magnitudes (\citealt{1996AJ....111.1748F}) to the luminosities using a standard $\Lambda$CDM cosmological model with
$(H_0, \Omega_M, \Omega_{\rm vac}, \Omega_k)=(70~{\rm{km~s^{-1} Mpc^{-1}}}, 0.3, 0.7, 0.0)$. 
Each of the five luminosities was then converted to any of the three
monochromatic luminosities at 5100\AA, 3000\AA, and 1350\AA\ using prescription of \cite{2015AcA....65..251K}. 
The monochromatic luminosity is calculated as a weighted mean of up to five luminosity values from the five SDSS filters, 
where as weights we used the sum of squares of the luminosity uncertainty and the conversion dispersion.
We provide 39,005 monochromatic luminosity estimates at 5100\AA, 139,686 at 3000\AA, and 229,248 at 1350\AA.

We have also downloaded the DR7Q data set, presented in \cite{2011ApJS..194...45S}, that contains the monochromatic luminosities and BH masses obtained directly from spectra.
We matched our derived monochromatic luminosities to the DR7Q data set.
In Figure~\ref{fig:Lum}, we provide comparisons for the three luminosities. In all three cases there is a tight correlation between the 
values obtained from the spectra and the broad-band magnitudes with minute mean offsets ($\sim$0.01 dex) and dispersions of 0.11, 0.10, 0.12 dex,
for 5100\AA, 3000\AA, and 1350\AA, respectively. 
Because our typical AGN is bright $\log{(L_{\rm mono}/\rm erg~s^{-1})}\approx 46$ and distant $z\approx2$, the
AGN host contamination is not or weakly present in this dataset. A fraction of the dispersion in the luminosity differences is caused by the AGN variability itself,
where we know that on time scales of months--years the flux can change by a 0.2--0.3 magnitude ($\sim$0.1 dex in luminosity) in optical bands 
(e.g., \citealt{1997ARA&A..35..445U,2010ApJ...721.1014M,2010ApJ...714.1194S,2016ApJ...826..118K}).

The bolometric luminosity $L_{\rm bol}$ is derived from the monochromatic luminosities at
5100\AA, 3000\AA, and 1350\AA\ using
the following bolometric corrections from \cite{2006ApJS..166..470R}: 9.26, 5.15, and 3.81 
respectively. By analogy to the monochromatic luminosity calculation, as the final bolometric luminosity
 we provide the weighted mean of the bolometric luminosities from the monochromatic luminosities, where the 
weights are the squared luminosity uncertainties. The Eddington luminosity $L_{\rm Edd}$ can be simply 
obtained from $L_{\rm Edd}=1.26 \times 10^{38}(M_{\rm BH}/M_\odot)$~erg~s$^{-1}$, 
where we first use the BH mass estimate based on the MgII line and if not present on CIV (derived and discussed below). 
The Eddington ratio is $\eta_{\rm Edd}=L_{\rm bol} L_{\rm Edd}^{-1}$.

\subsection{The Black Hole Mass}

The virial BH mass is typically obtained from a single spectrum using
\begin{equation}
\log\left(\frac{M_{\rm BH}}{M_\odot}\right)=a+b\log\left(\frac{L\times 10^{-44}}{{\rm erg~s}^{-1}}\right) + 2\log\left(\frac{\rm FWHM}{{\rm km~s}^{-1}}\right),
\end{equation}
where the BH mass calibrations $(a,b)$ are estimated empirically (e.g., \citealt{2002MNRAS.331..795M,2006ApJ...641..689V,2011ApJS..194...45S}), typically against nearby reverberation mapped AGNs 
(e.g., \citealt{1982ApJ...255..419B,1993PASP..105..247P,2004ApJ...613..682P}).

\begin{figure*}
\centering
\includegraphics[width=7.0cm]{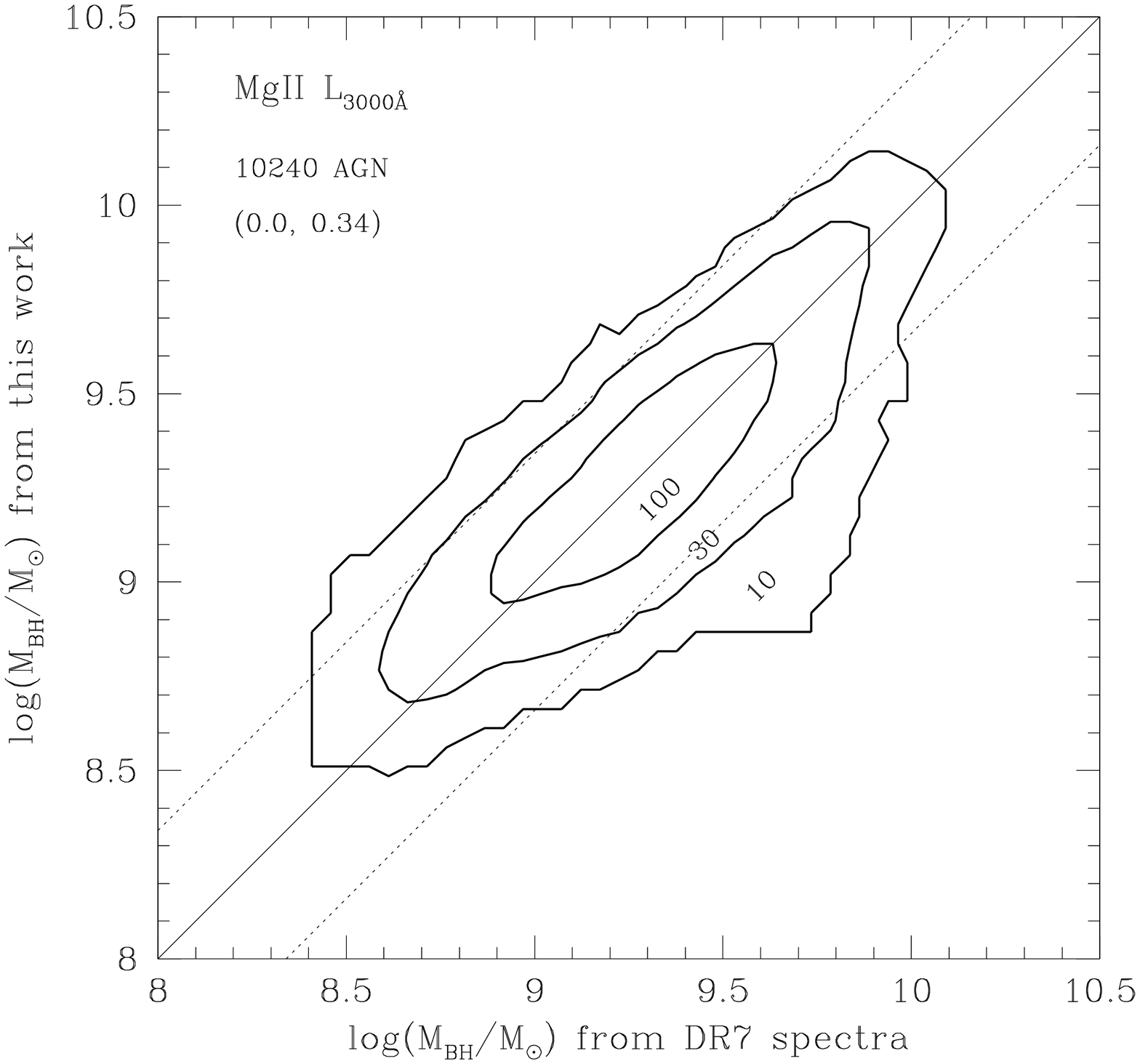}
\includegraphics[width=7.0cm]{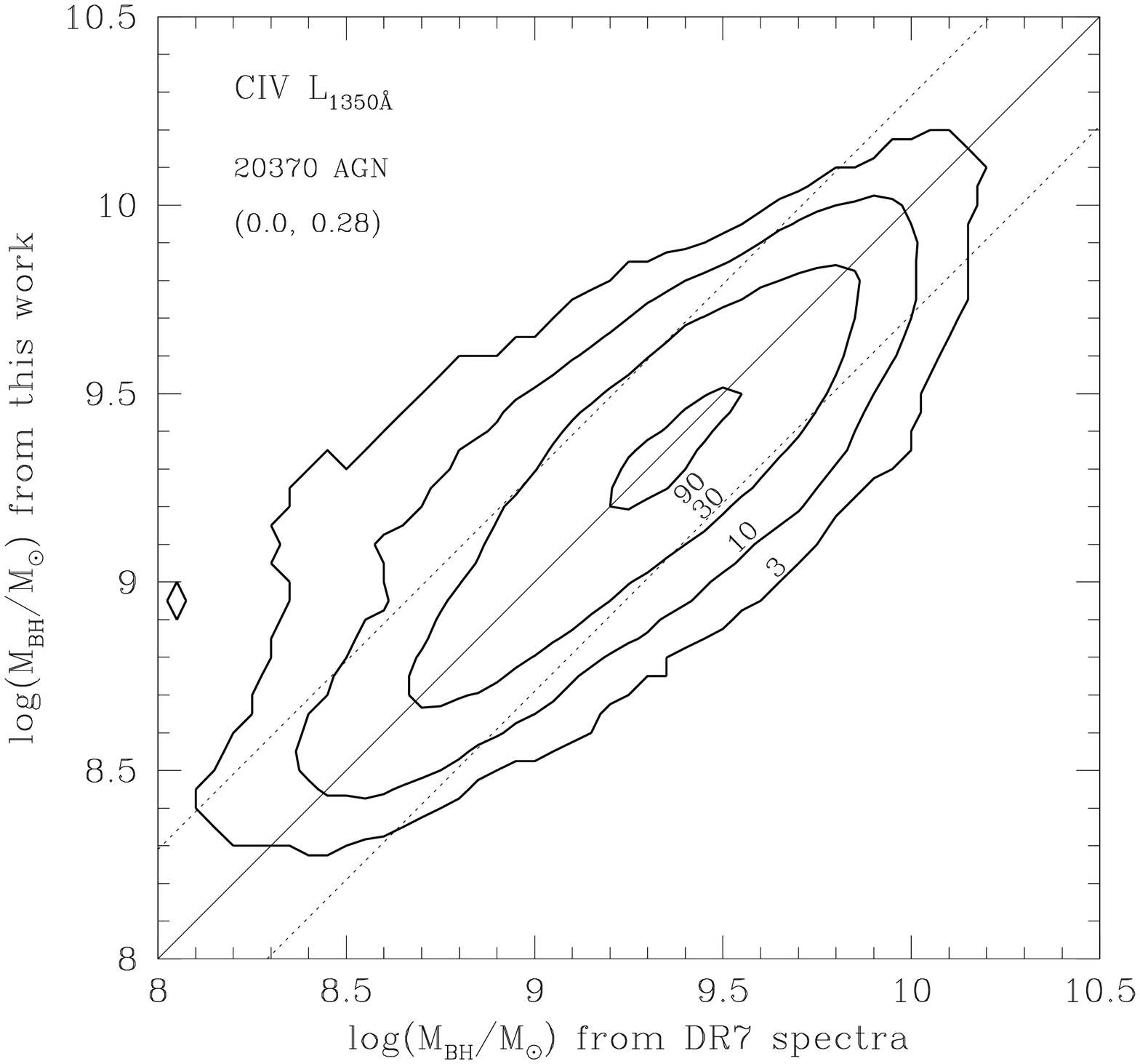}
\caption{Comparison of the BH masses from DR7Q (x-axis) and calculated from the broad-band magnitudes (y-axis) for the MgII line (left panel) and the CIV line (right panel). 
Contours show the number of objects per $0.05 \times 0.05$ dex bin. In brackets, we provide the mean and dispersion of differences, respectively.}
\label{fig:Mcomp}
\end{figure*}

{\it Masses based on the MgII line:} The coefficients for MgII are $(a,b)=(0.74,0.62)$ from \cite{2011ApJS..194...45S} and produce an offset of 0.056 dex in the derived masses
as compared to the matched DR7Q BH masses. We, therefore, modify them to $(a,b)=(0.796,0.62)$. 
The same pair of values is found by the minimization of the BH mass differences in a least squares sense on an $(a,b)$ grid.
In the left panel of Figure~\ref{fig:Mcomp}, we present the comparison between the BH masses derived from the SDSS spectra (DR7Q; x-axis)
and the ones derived from this work (y-axis). The mean difference offset is by definition 0.0 dex, because the $(a,b)$ parameters are found on a grid,
while the dispersion of differences is 0.34 dex.

\begin{figure}
\centering
\includegraphics[width=7.0cm]{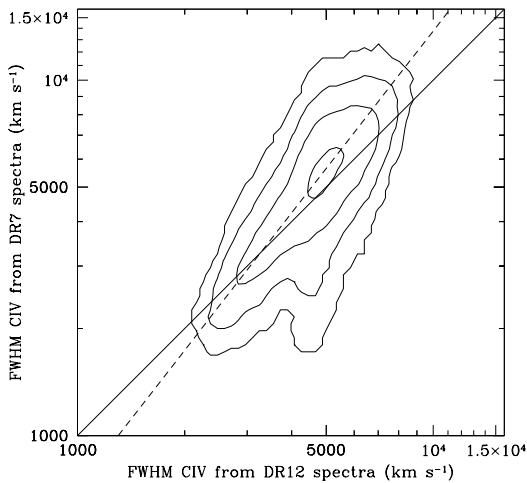}
\caption{Bias in the FWHM of the CIV line between the DR7Q and DR12Q. While the bulk of population should be located along the solid line, it is obvious
that there exist a bias between the two measurement methods. We correct the bias by fitting the linear trend to the bulk of the data
marked by the dashed line and Eq.~(\ref{fig:bias}).
Contours show the density of objects per $0.05 \times 0.05$ dex bin.}
\label{fig:FWHM}
\end{figure}

{\it Masses based on the CIV line:}
For the CIV line, \cite{2011ApJS..194...45S} use $(a,b)=(0.66,0.53)$ from \cite{2006ApJ...641..689V},
but we found a rather biased BH masses using DR12Q as compared to the DR7Q ones, and to the ones based on the MgII line. 
We found that while the MgII FWHMs in DR7Q and DR12Q match one another, there is a problem with the FWHMs of CIV (Figure~\ref{fig:FWHM}). 
We use the following empirically found conversion to match DR12Q CIV FWHMs to the DR7Q ones

\begin{equation}
\log\left(\frac{\rm FWHM_{7}}{{\rm km~s}^{-1}}\right)=3.62 + 1.286\left(\log\left(\frac{\rm FWHM_{12}}{{\rm km~s}^{-1}}\right)-3.6\right).
\label{fig:bias}
\end{equation}

After exploring the reason of to why the two data releases differ in this context, which to some degree is beyond the scope of this paper,
we find that DR7Q FWHMs appear to be more reliable than the DR12Q ones. This is because \cite{2011ApJS..194...45S} carefully fits to spectra all the necessary 
ingredients (simultaneously: the continuum, the Fe template, broad and narrow Gaussians for the lines) to measure the FWHMs that are in fact 
converted from the Gaussian dispersion. And the Gaussian dispersion is a more robust estimate of the BLR velocity than
a straight FWHM measurement for the CIV line (\citealt{2013ApJ...775...60D}). \cite{2016arXiv160806483P}, on the other hand, use
principal component analysis to obtain the reported FWHMs. The CIV line is also known to be notoriously troubled by other effects such
as a significant blue shift of up to thousands of km~s$^{-1}$ (e.g., \citealt{2002AJ....124....1R}) and 
the line asymmetry due to outflows (e.g., \citealt{1982ApJ...263...79G}) 
that seem not to reverberate (\citealt{2012ApJ...759...44D}), blurring the picture of the BH mass estimation.

From a minimization of differences between our BH masses and the ones from DR7Q, that we perform 
on an $(a,b)$ grid, we find $(a,b)=(0.64,0.53)$ to be the best choice for the CIV BH masses.
In the right panel of Figure~\ref{fig:Mcomp}, we present the comparison between the CIV BH masses derived from the SDSS spectra (DR7Q; x-axis)
and the ones derived from this work after correcting the FWHMs from DR12Q (y-axis). The mean offset by definition is 0.0 dex, while the dispersion
is 0.28 dex. Because the CIV line is known for being problematic and/or biased in terms of the FWHM measurement (e.g., \citealt{2008ApJ...680..169S}),
we caution the user of this BH mass estimate.

\begin{figure*}
\centering
\includegraphics[width=7.0cm]{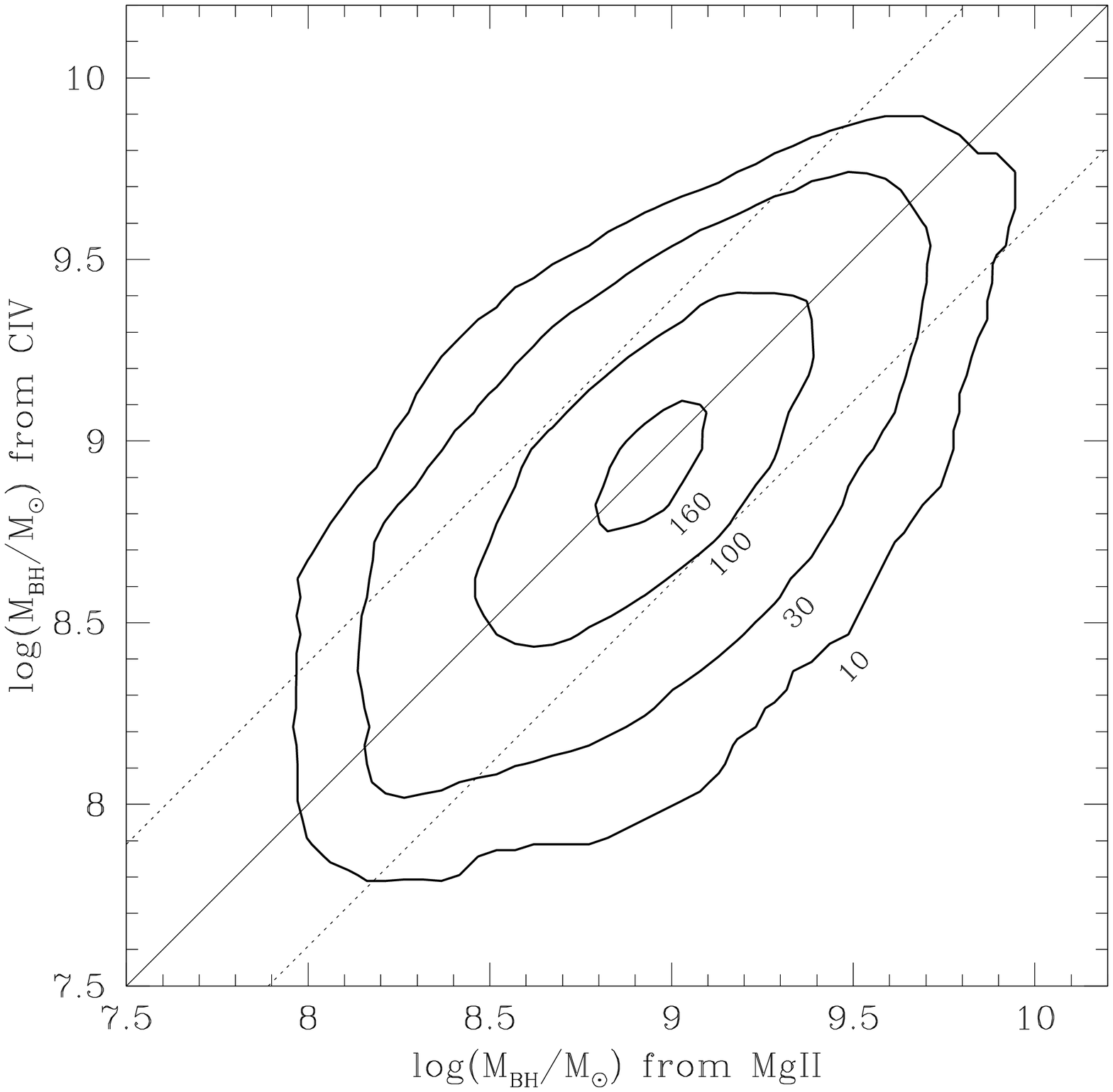}
\includegraphics[width=7.0cm]{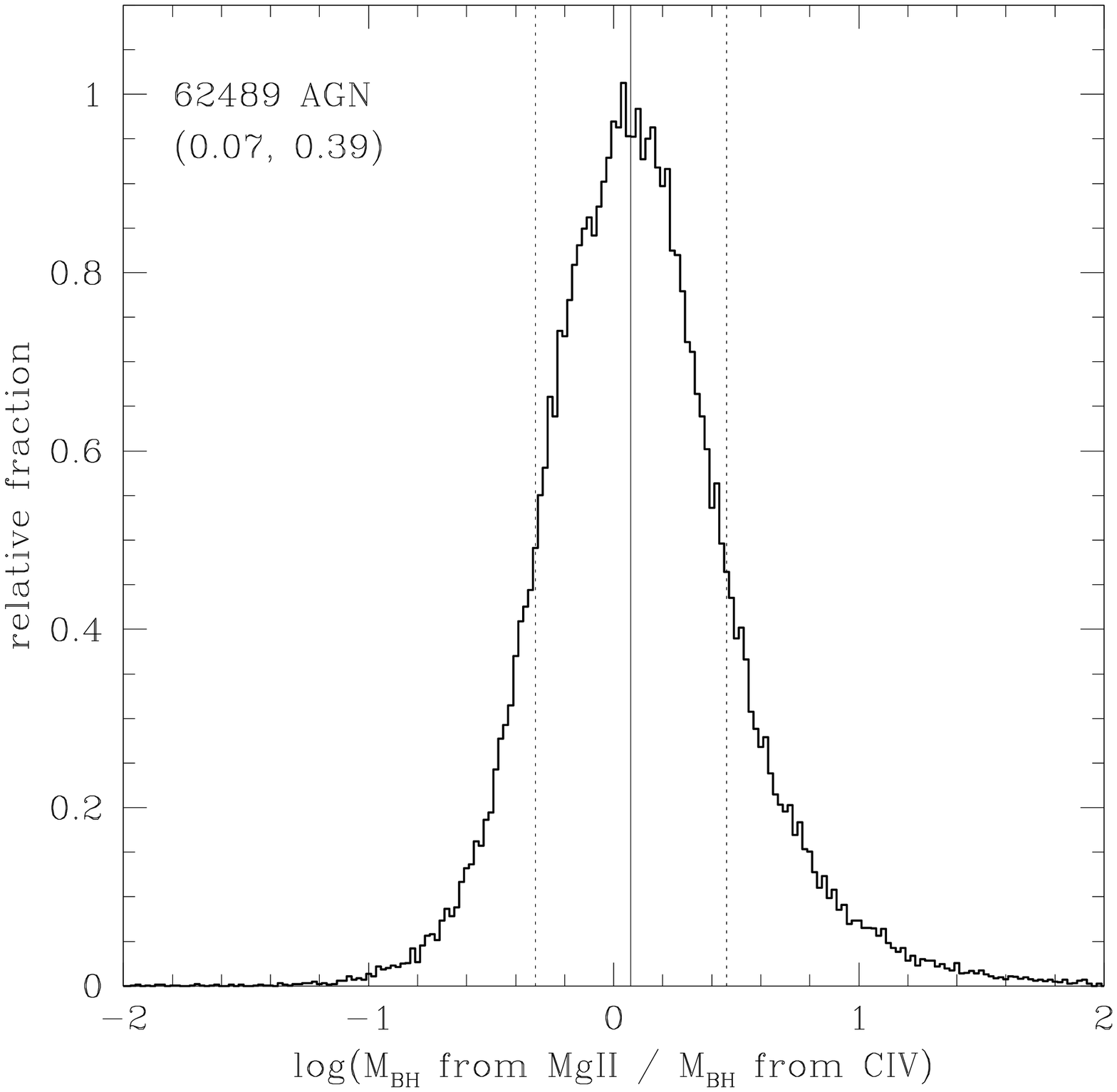}
\caption{Left panel: Internal comparison of the MgII (x-axis) and CIV (y-axis) BH masses derived from the broad-band SDSS magnitudes (converted to the monochromatic luminosity).
Contours show the number of objects per $0.05 \times 0.05$ dex bin.
Right panel: The normalized histogram of ratios of the MgII to CIV BH masses is shown. 
In brackets we provide the mean and dispersion of differences, respectively.}
\label{fig:MBH}
\end{figure*}

We also perform the internal comparison between the MgII and CIV BH masses based on DR12Q.
In the left panel of Figure~\ref{fig:MBH}, we present the density contours, and in the right panel a histogram of differences of BH mass logarithms.
We find a small offset between the two samples of 0.07 dex (MgII BH masses are larger) and the dispersion of 0.39 dex after pruning 3$\sigma$ outliers.

The monochromatic and bolometric luminosities, the Eddington ratio, and the BH mass estimates are provided in Table~\ref{tab:params}.

\begin{deluxetable*}{lrcc|cccccc|cc|ccc}
\tablecaption{Estimates of the Monochromatic and Bolometric Luminosities, Eddington Ratios, and Black Hole Masses for DR12Q AGNs.\label{tab:params}}
\tablewidth{0pt}
\scriptsize
\tablehead{
RA & Decl. & z & $M_i$ & $L_{5100}$ & $\sigma_{L_{5100}}$ & $L_{3000}$ & $\sigma_{L_{3000}}$ & $L_{1350}$ & $\sigma_{L_{1350}}$ & \multicolumn{2}{c}{$M_{\rm BH}$ from} & $L_{\rm bol}$ & $\sigma_{L_{\rm bol}}$ & $\eta_{\rm Edd}$\\
(deg) & (deg) & & (mag) & & & & & & & MgII & CIV & & & }
\startdata
0.00806 &  $-$0.24097 & 2.16 & $-25.79$ & $-99.99$ & $-9.99$ & 45.63 & 0.11 & 45.83 & 0.10 & 9.25 & 9.39 & 46.39 & 0.076 & $-0.96$ \\
0.00855 &    34.67868 & 1.57 & $-23.20$ & $-99.99$ & $-9.99$ & 44.66 & 0.09 & 44.83 & 0.11 & 8.09 & 8.39 & 45.39 & 0.070 & $-0.80$ \\
0.00898 &    15.25463 & 1.71 & $-24.56$ & $-99.99$ & $-9.99$ & 45.04 & 0.08 & 45.12 & 0.13 & 8.72 & 8.27 & 45.74 & 0.070 & $-1.07$ \\
0.01471 &    30.57038 & 2.16 & $-24.44$ & $-99.99$ & $-9.99$ & 45.04 & 0.12 & 45.22 & 0.10 & 8.32 & 8.51 & 45.78 & 0.082 & $-0.63$ \\
0.01644 &    26.61267 & 2.18 & $-26.42$ & $-99.99$ & $-9.99$ & 45.86 & 0.12 & 46.07 & 0.09 & 9.19 & 9.10 & 46.62 & 0.074 & $-0.66$
\enddata
\tablecomments{One digit has been truncated from each column in this illustrative table in order to fit the page;
$L_{5100}$, $\sigma_{L_{5100}}$, $L_{3000}$, $\sigma_{L_{3000}}$, $L_{1350}$, $\sigma_{L_{1350}}$, 
and $L_{\rm bol}$ are the base 10 logarithms of the luminosities and their uncertainty (at 5100\AA, 3000\AA, 1350\AA, and bolometric) 
in units of erg~s$^{-1}$; $M_{\rm BH}$ are the base 10 logarithms of the BH mass in units of $M_\odot$ calculated
via the MgII or CIV line; $\eta_{\rm Edd}$ is the base 10 logarithm of the Eddington ratio.
The error code, reflecting no measurement, is $-99.99$ ($-9.99$ for the uncertainty).
(This table is available in its entirety in a machine-readable form in the online journal. A portion is shown here for guidance regarding its form and content.
The data file containing the presented results can be also obtained from: \tt ftp://ftp.astrouw.edu.pl/pub/simkoz/SDSS-DR12Q-BH.tar.gz)}
\end{deluxetable*}

%%%%%%%%%%%%%%%%%%%%%  DISCUSSION  %%%%%%%%%%%%%%%%%%%%

\section{Discussion}
\label{sec:discussion}

While \cite{2002MNRAS.337..109M} and \cite{2011ApJ...742...93A} find that CIV FWHM line measurements are generally consistent with those for hydrogen lines, 
\cite{2013ApJ...775...60D}, \cite{2016ApJS..224...14D}, and \cite{2016arXiv160508057D} find that the decreasing signal-to-noise ratio can significantly bias the measured CIV FWHMs and therefore BH masses.
In particular, \cite{2013ApJ...775...60D} show that for the high signal-to-noise spectra the line dispersion, rather than FWHM, returns reliable BH estimates, and
 \cite{2016arXiv160508057D} identify several problems in the CIV FWHM measurements due to (1) intrinsic diversity of AGNs and 
(2) systematic offsets between the CIV and HeII or [OII]-based redshifts, leading to the reported biased redshifts from the SDSS/BOSS pipeline (used in DR12Q).
The latter is studied in \cite{2016arXiv160203894S}, who point out problems with the systemic velocity shifts in AGNs and their impact on the measured redshifts. 
While this may look as a weakly important issue here, incorrectly measured redshifts can lead to biases in the derived luminosities, what in turn leads 
to biases in the BH masses.

Table~\ref{tab:params} presents our BH mass estimates. It is important to understand that the initial 
DR7Q BH mass estimates are based on the luminosity-BLR radius relation that have
intrinsic dispersion of 0.4 dex (e.g., see \citealt{2013BASI...41...61S} for a review). Because \cite{2016arXiv160806483P} do not provide uncertainties for their FWHMs,
we are dealing with a 0.4 dex systematic bias to start with, and we have only estimates on the uncertainties of the luminosity, we have decided to report 
estimates of the BH masses only (without uncertainties). These masses should serve as a means for statistical studies of large samples of AGNs, 
keeping in mind a plausible systematic $\sim$0.4 dex offsets, rather than for studying individual objects.

%%%%%%%%%%%%%%%%%%%%%  SUMMARY  %%%%%%%%%%%%%%%%%%%%

\section{Summary}
\label{sec:summary}

In this paper, we used the twelfth quasar data release from SDSS to obtain basic physical parameters for $\sim$280,000 AGNs,
that include black hole masses, luminosity at 5100\AA, 3000\AA, 1350\AA, and bolometric, and the Eddington ratio.
First, we estimated the monochromatic luminosities from the broad-band extinction-corrected $ugriz$ SDSS magnitudes. They were then matched to and compared
to the ones derived from spectra and published in DR7Q. We find excellent correlations with the mean difference offsets of $\sim$0.01 dex
and dispersion of $\sim$0.1 dex. Then, by combining these luminosities with the broad emission line widths, we estimate the black hole masses.
For the MgII line we find an excellent match to the DR7Q BH masses, while for the CIV line we uncover a bias between the two data releases: DR7Q and DR12Q,
due to the method used to estimate FWHMs. Nevertheless, we correct this bias empirically, after which our CIV BH masses match nicely 
with the ones from DR7Q. We also cross-check the MgII and CIV BH masses internally, and we find a mean offset of differences of 0.07 dex with 
the dispersion of 0.39 dex. All the obtained physical parameters of AGNs are provided in a tabular form line-by-line matched to the DR12Q catalog from \cite{2016arXiv160806483P}.
There are 283,033 virial black hole mass estimates, of which 132,451 are for MgII, 213,071 are for CIV, and 62,489 are for the both lines.

While there are many (some unknown) uncertainties in the quest for obtaining the BH masses (the intrinsic 0.4 dex dispersion in the BLR radius-luminosity relation, unknown
uncertainties of FWHMs from DR12Q), the estimates provided in this work are better suited for statistical analyses of AGNs, in particular, they could be used in 
tracing correlations of the AGN variability (from large ground-based and/or space-based sky surveys) with the physical AGN parameters. 
Because the CIV line is typically problematic to correctly measure 
from spectra (both its width and centroid), is often asymmetric, and a fraction of light is unrelated to the bulk velocity motion of BLR (it does not reverberate),
we caution the reader that the CIV line-based BH masses are simply and only best estimates from the available DR12Q catalog.

%%%%%%%%%%%%%%%%%%%%%  ACKNOWLEDGMENTS  %%%%%%%%%%%%%%%%%%%%

\acknowledgments
I am grateful to Drs. Jan Skowron, {\L}ukasz Wyrzykowski, and Pawe{\l} Pietrukowicz for discussions of this topic.
This work has been supported by the Polish National Science Centre grant No. DEC-2014/15/B/ST9/00093.

%%%%%%%%%%%%%%%%%%%%%  BIBLIOGRAPHY  %%%%%%%%%%%%%%%%%%%%


\begin{thebibliography}{}

\bibitem[Assef et al.(2011)]{2011ApJ...742...93A} Assef, R.~J., Denney, K.~D., Kochanek, C.~S., et al.\ 2011, \apj, 742, 93 

\bibitem[Bentz et al.(2009)]{2009ApJ...697..160B} Bentz, M.~C., Peterson, B.~M., Netzer, H., Pogge, R.~W., \& Vestergaard, M.\ 2009, \apj, 697, 160 

\bibitem[Blandford \& McKee(1982)]{1982ApJ...255..419B} Blandford, R.~D., \& McKee, C.~F.\ 1982, \apj, 255, 419 

\bibitem[Denney(2012)]{2012ApJ...759...44D} Denney, K.~D.\ 2012, \apj, 759, 44

\bibitem[Denney et al.(2013)]{2013ApJ...775...60D} Denney, K.~D., Pogge, R.~W., Assef, R.~J., et al.\ 2013, \apj, 775, 60 

\bibitem[Denney et al.(2016a)]{2016ApJS..224...14D} Denney, K.~D., Horne, K., Shen, Y., et al.\ 2016a, \apjs, 224, 14

\bibitem[Denney et al.(2016b)]{2016arXiv160508057D} Denney, K.~D., Horne, K., Brandt, W.~N., et al.\ 2016b, arXiv:1605.08057

\bibitem[Fukugita et al.(1996)]{1996AJ....111.1748F} Fukugita, M., Ichikawa, T., Gunn, J.~E., et al.\ 1996, \aj, 111, 1748 

\bibitem[Gaskell(1982)]{1982ApJ...263...79G} Gaskell, C.~M.\ 1982, \apj, 263, 79 

\bibitem[Kaspi et al.(2000)]{2000ApJ...533..631K} Kaspi, S., Smith, P.~S., Netzer, H., et al.\ 2000, \apj, 533, 631 

\bibitem[Kaspi et al.(2007)]{2007ApJ...659..997K} Kaspi, S., Brandt, W.~N., Maoz, D., et al.\ 2007, \apj, 659, 997 

\bibitem[Koz{\l}owski(2015)]{2015AcA....65..251K} Koz{\l}owski, S.\ 2015, \actaa, 65, 251 

\bibitem[Koz{\l}owski(2016)]{2016ApJ...826..118K} Koz{\l}owski, S.\ 2016, \apj, 826, 118

\bibitem[MacLeod et al.(2010)]{2010ApJ...721.1014M} MacLeod, C.~L., Ivezi{\'c}, {\v Z}., Kochanek, C.~S., et al.\ 2010, \apj, 721, 1014 

\bibitem[McLure \& Dunlop(2002)]{2002MNRAS.331..795M} McLure, R.~J., \& Dunlop, J.~S.\ 2002, \mnras, 331, 795 

\bibitem[McLure \& Jarvis(2002)]{2002MNRAS.337..109M} McLure, R.~J., \& Jarvis, M.~J.\ 2002, \mnras, 337, 109 

\bibitem[P{\^a}ris et al.(2016)]{2016arXiv160806483P} P{\^a}ris, I., Petitjean, P., Ross, N.~P., et al.\ 2016, arXiv:1608.06483 

\bibitem[Peterson(1993)]{1993PASP..105..247P} Peterson, B.~M.\ 1993, \pasp, 105, 247 

\bibitem[Peterson et al.(2004)]{2004ApJ...613..682P} Peterson, B.~M., Ferrarese, L., Gilbert, K.~M., et al.\ 2004, \apj, 613, 682

\bibitem[Richards et al.(2002)]{2002AJ....124....1R} Richards, G.~T., Vanden Berk, D.~E., Reichard, T.~A., et al.\ 2002, \aj, 124, 1 

\bibitem[Richards et al.(2006)]{2006ApJS..166..470R} Richards, G.~T., Lacy, M., Storrie-Lombardi, L.~J., et al.\ 2006, \apjs, 166, 470 

\bibitem[Salpeter(1964)]{1964ApJ...140..796S} Salpeter, E.~E.\ 1964, \apj, 140, 796 

\bibitem[Schlafly \& Finkbeiner(2011)]{2011ApJ...737..103S} Schlafly, E.~F., \& Finkbeiner, D.~P.\ 2011, \apj, 737, 103 

\bibitem[Schmidt et al.(2010)]{2010ApJ...714.1194S} Schmidt, K.~B., Marshall, P.~J., Rix, H.-W., et al.\ 2010, \apj, 714, 1194

\bibitem[Shakura \& Sunyaev(1973)]{1973A&A....24..337S} Shakura, N.~I., \& Sunyaev, R.~A.\ 1973, \aap, 24, 337

\bibitem[Shen et al.(2008)]{2008ApJ...680..169S} Shen, Y., Greene, J.~E., Strauss, M.~A., Richards, G.~T., \& Schneider, D.~P.\ 2008, \apj, 680, 169-190 

\bibitem[Shen et al.(2011)]{2011ApJS..194...45S} Shen, Y., Richards, G.~T., Strauss, M.~A., et al.\ 2011, \apjs, 194, 45 

\bibitem[Shen(2013)]{2013BASI...41...61S} Shen, Y.\ 2013, Bulletin of the Astronomical Society of India, 41, 61

\bibitem[Shen et al.(2015)]{2015ApJS..216....4S} Shen, Y., Brandt, W.~N., Dawson, K.~S., et al.\ 2015, \apjs, 216, 4 

\bibitem[Shen et al.(2016)]{2016arXiv160203894S} Shen, Y., Brandt, W.~N., Denney, K.~D., et al.\ 2016, arXiv:1602.03894 

\bibitem[Ulrich et al.(1997)]{1997ARA&A..35..445U} Ulrich, M.-H., Maraschi, L., \& Urry, C.~M.\ 1997, \araa, 35, 445

\bibitem[Vestergaard \& Peterson(2006)]{2006ApJ...641..689V} Vestergaard, M., \& Peterson, B.~M.\ 2006, \apj, 641, 689 

\end{thebibliography}
\end{document}